# BEAM ENERGY STABILIZATION OF THE KEK 40MEV PROTON LINAC

Z.Igarashi, K.Nanmo, T.Takenaka and E.Takasaki

High Energy Accelerator Research Organization, KEK

Oho1-1, Tsukuba, Ibaraki, 305-0801, Japan

*Abstract*

The new method to stabilize the beam energy of the KEK 40MeV proton linac, is developed now. In this method, the signal of the velocity monitor installed upstream the debuncher in the 40MeV beam line, is processed and then fed to the phase shifter of the debuncher rf system so as to cancel the fluctuation of the beam energy.

In this article, the beam tests to prove the validity of this method and the system are described.

## 1 INTRODUCTION

Various improvements to increase the beam intensity of the KEK 12GeV PS for neutrino oscillation experiments have been continued these several years[1]. Since one of the causes that limit the beam intensity is the beam loss in the accelerators or the beam lines, the accelerators should be tuned with scrupulous care and these conditions be kept constant during the long operation period.

One of the beam parameters that affect the next accelerations and the beam extractions is the centre energy (momentum) of the beam. In order to keep the beam energy being constant in the proton linac, the rf system should equip the feedback loop which stabilizes the accelerating field. Unfortunately, since the rf system of our linac is operated near the saturation, the effects of the feedback are not expected much. Hence, the investigation of the new method to stabilize the beam energy was started.

## 2 BEAM TESTS

### 2.1 The layout of the linac

The KEK 40MeV proton linac that consists of the prebuncher, the 20MeV tank, the 40MeV tank, the debuncher, and the 40MeV beam line is shown in Figure 1. Two velocity ($\beta$) monitors are installed in the 40MeV beam line, one ($40\beta 1$) is for the detection of the output beam of the linac and the other ($40\beta 2$) is for the beam to inject the 500MeV booster synchrotron. The debuncher is installed between them[2],[3]. The typical waveforms of the $20\beta 1$, the $40\beta 1$ and the $40\beta 2$ are shown in Figure 2. Figure 3 shows the variations of the $40\beta 1$ and the linac beam current during the operation period of about a month. The fluctuations of the accelerating energy are within 0.9%.

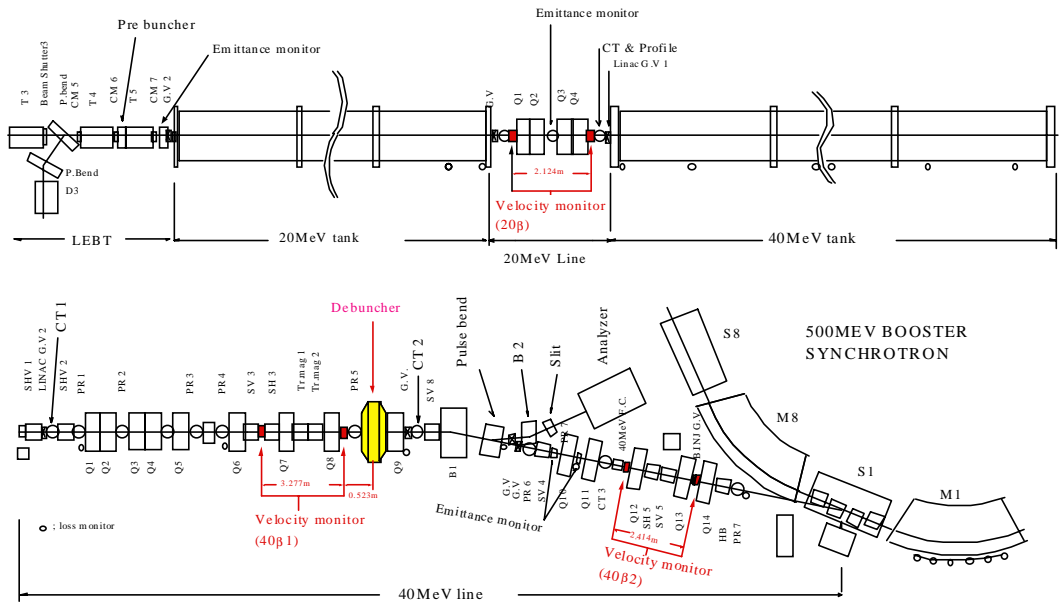

Figure 1: The layout of the KEK 40MeV proton linac and the beam lines.

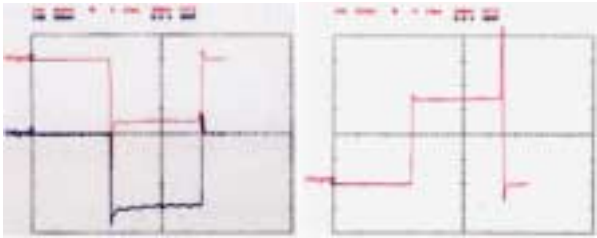

Figure 2: 20β(left upper), 40β1(left lower) and 40β2(right)

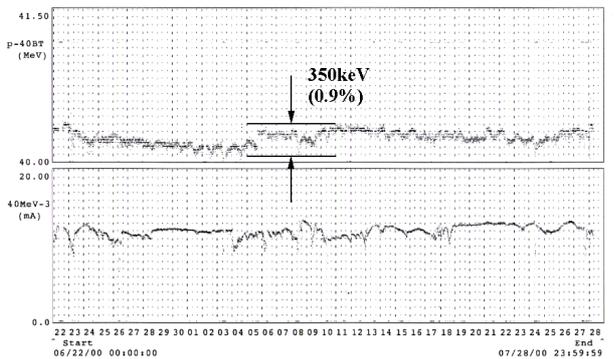

Figure 3: The variations of the 40β1(upper) and the linac beam current(lower).

## 2.2 Beam tests

Figure 4 shows the layout of the beam tests. The velocity monitor signals are acquired by the VME system and simultaneously observed by the scope. The debuncher rf system, therefore, the phase shifter is controlled by the PLC(Programmable Logic Controller).

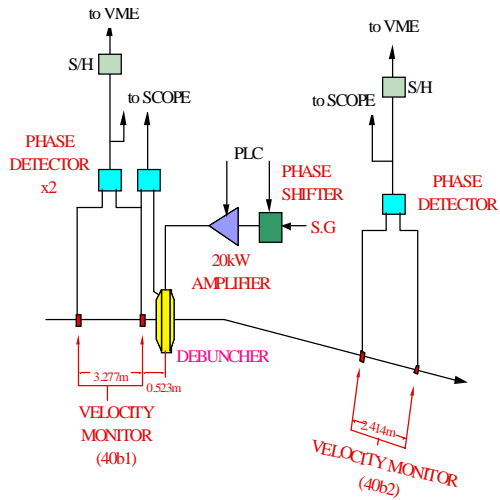

Figure 4: The layout of the beam tests.

The result of the energy variation measured by 40β1 and 40β2 versus the debuncher phase is shown in Figure 5. Though the variation of the momentum spread (ΔP/P), it is obvious from Figure 5 that it is possible to change the beam energy within ±300MeV.

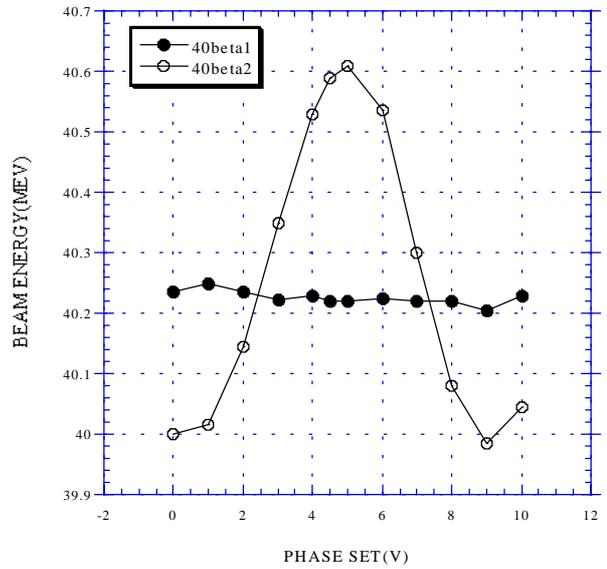

Figure 5: Energy variations versus the debuncher phase.

In order to estimate the new method, we studied whether it cancels the energy fluctuations due to the accelerating field of the two tanks, namely, the 20MeV tank level, the 40MeV tank level and the phase between two tanks.

Figure 6, Figure 7 and Figure 8 are the results of these tests. In these graphs, the 40beta1 is the plots for the linac output energy, the 40beta2@D.B ON for the constant phase of the debuncher and the 40beta2@P.ADJ for the adjusted phase to cancel the energy fluctuation.

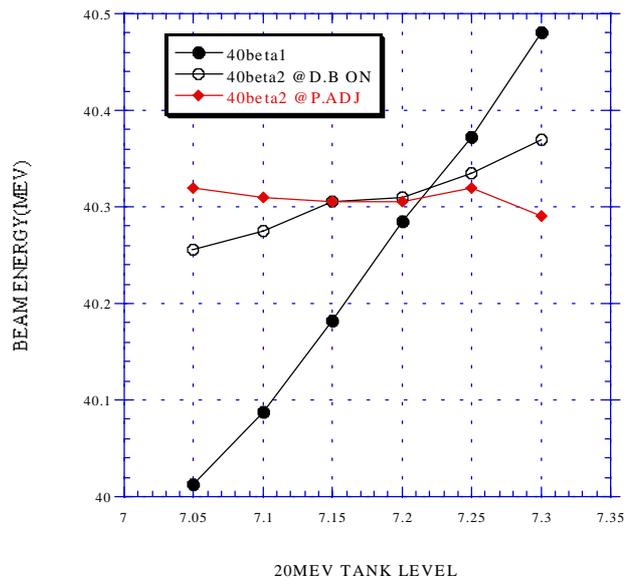

Figure 6: The effect of the fluctuation of the 20MeV tank level

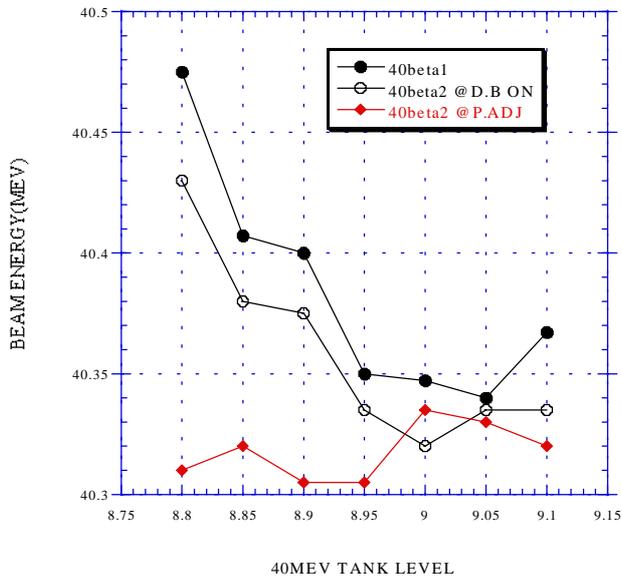

Figure 7: The effect of the fluctuation of the 40MeV tank level.

In all tests, the input rf power to the debuncher is 15.5kW.
From these results, the fluctuations due to the accelerating fields of the linac are within 0.08%.

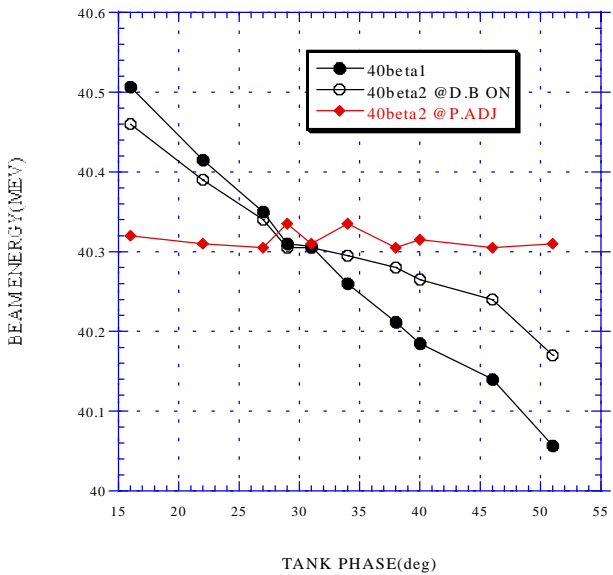

Figure 8: The effect of the fluctuation of the tank phase.

## 3 THE SYSTEM PLAN FOR THIS METHOD

In this method, to stabilize the beam energy, the debuncher phase should be set to the appropriate value that is caluculated from the signals of the velocity monitor and the phasing system between the beam and the debuncher field. Furthermore, for the stabilization during the pulse duration of the beam, the WE 7000 SYSTEM made by YOKOGAWA ELECTRIC COMPANY will be introduced as the data acquisition and the control system.

## 3 SUMMARY

It is proved that the new method by using the velocity monitors and the debuncher is effective to stabilize the beam energy. Especially, the energy fluctuation due to the 20MeV tank level, the 40MeV tank level and the phase between two tanks are reduced within 0.08%.

It is expected that the new system will be completed immediately and then will be used for the normal operation of the KEK 12GeV PS.

## REFERENCE

[1] I.Yamane, H.Sato "Accelerator Development for K2K Long-Baseline Neutrino-Oscillation Experiment," January, 2000
[2] Z.Igarashi, K.Nanmo, T.Takenaka and E.Takasaki,"Velocity Monitor for the KEK 40MeV Proton Linac," Proc. 1992 Linac Conference., (1992)
[3] Z.Igarashi, K.Nanmo, T.Takenaka and E.Takasaki, "A New RF System for the Debuncher at the KEK 40-Mev Proton Linac," Proc. 1998 Linac Conference., 929 (1998)